\documentclass[final,5p,times,twocolumn,authoryear]{elsarticle}

\usepackage{amssymb}
\usepackage{amsmath}
\usepackage{lipsum}
\usepackage{float}
\usepackage{threeparttable}

\journal{High Energy Astrophysics}

\begin{document}

\begin{frontmatter}



\title{\bf Probing the nova shock physics with future gamma-ray observations of the upcoming outburst from T Coronae Borealis}

\affiliation[1]{organization={School of Astronomy and Space Science, Nanjing University},
            city={Nanjing},
            postcode={210093}, 
            country={China}}
\affiliation[2]{organization={Key Laboratory of Modern Astronomy and Astrophysics (Nanjing University)},
            city={Nanjing},
            postcode={210093}, 
            country={China}}
\affiliation[3]{organization={Institute of High Energy Physics, Chinese Academy of Sciences},
            city={Beijing},
            postcode={100049}, 
            country={China}}
\affiliation[4]{organization={University of Chinese Academy of Sciences},
            city={Beijing},
            postcode={100049}, 
            country={China}}
\affiliation[5] {organization=  {Tianfu Cosmic Ray Research Center},
            city={Chengdu},
            postcode={610000}, 
            country={China}}
\author[1,2]{Jian-He Zheng}
\author[1,2]{Hai-Ming Zhang}
\author[1,2,5]{Ruo-Yu Liu}
\author[3,4,5]{Min Zha}
\author[1,2,5]{Xiang-Yu Wang}
\ead{xywang@nju.edu.cn}

\begin{abstract}
Nova shocks behave like scaled-down supernova remnant shocks with a lifetime of only a few weeks or months, thereby providing a unique opportunity to study the dynamics of non-relativistic shocks as well as the shock acceleration physics.
Recently, GeV and TeV gamma-ray emissions from an outburst of the recurrent nova RS~Ophiuchi have been observed. The light curves of the gamma-ray emissions suggest that they arise from an external shock, which is formed as the nova ejecta interacts with the ambient medium. The shock is thought to transition from an adiabatic shock to a radiative one at later times, but no such later observations are available for RS~Ophiuchi. In addition, the spectral evolution of the gamma-ray outburst of RS~Ophiuchi was not well measured, and hence the related particle acceleration mechanisms are not well understood. T Coronae Borealis (T~CrB) is another recurrent nova in Milky Way and its last outburst was nearly ten times optically brighter than RS~Ophiuchi. Recently the optical light curve of T~CrB  displayed a state transition behavior before the eruption, and it has been predicted that T~CrB will undergo an outburst in the near future. By performing a theoretical investigation, we find that Fermi-LAT could probably capture the transition of the shock from the adiabatic phase to the radiative phase at the GeV band if the ambient wind medium is dense with $A_{\star}\geq1$, thanks to a longer detectable time than that of RS~Ophiuchi. 
Due to its higher brightness, we also find that imaging atmospheric Cherenkov telescopes (IACTs) such as MAGIC and VERITAS, and extensive air shower experiments such as LHAASO could detect the nova outburst and measure the gamma-ray spectrum in the very-high-energy (VHE, $>0.1\,{\rm TeV}$) band more precisely. This can be used to constrain the high-energy cutoff index in the accelerated proton spectrum and the acceleration efficiency, which will shed light on the particle acceleration physics in nova shocks.

\end{abstract}



\begin{keyword}
nova \sep gamma-ray astronomy \sep cosmic-ray



\end{keyword}

\end{frontmatter}




\section{Introduction}
\label{sec:intro}
Novae are energetic explosions that take place in binary star systems in which white dwarfs (WD) accrete matter from their companion stars. Accreted materials accumulate on the surface of the WD, leading to the rise of the temperature, which finally results in a thermonuclear runaway (TNR) when the layer reaches the critical mass \citep[e.g.][]{Townsley2004ApJ...600..390T}. Novae are usually classified into two categories based on the nature of companion stars: classical novae and novae in the symbiotic systems. Classical novae are binary systems comprising of a main sequence star and a WD. Binary systems with a red giant (RG) companion and a WD are novae in symbiotic systems \citep{Chomiuk2021ARA&A..59..391C}.

After the launch of Fermi Large Area Telescope (LAT), a dozen of novae were discovered in the GeV gamma-ray band, confirming that nova outbursts can produce strong shocks and accelerate relativistic particles \citep[e.g.][]{fermi2010Sci...329..817A,Cheung2016ApJ...826..142C}. In classical novae, gamma-ray emissions are thought to arise from internal shocks that occur as a result of internal collisions between distinct components of the nova ejecta. So far, multiwavelength observations from radio to X-rays have shown that internal shocks are common in classical novae \citep[e.g.][]{Chomiuk2014Natur.514..339C,Nelson2019ApJ...872...86N}. The GeV light curves of classical novae usually exhibit clear fluctuations. 

On the other hand, if the companion star is a red giant star, the WD is embedded in a dense wind from the RG. Ejecta from nova outburst collides with the dense wind, generating an expanding external shock that accelerates relativistic particles. Unlike internal shocks, the light curve of the gamma-ray emission from external shocks is expected to be very smooth.

The first gamma-ray detected nova V407 Cygni (V407 Cyg) was found to be embedded in a stellar wind from a Mira-like red giant. In March, 2010, the gamma-ray emission of V407 Cyg was observed around the optical peak and lasted a half month \citep{fermi2010Sci...329..817A}. The radiation mechanism and the nature of the shocks are still debated. Non-thermal emission from protons and electrons can both explain the spectra of the gamma-ray emission \citep{fermi2010Sci...329..817A}. Although an external shock expanding in the dense wind is a natural explanation for V407 Cyg \citep[e.g.][]{Martin2013A&A...551A..37M}, the internal shock model is also suggested to be able to explain the gamma-ray data \citep{Martin2018A&A...612A..38M}.

In 2021, the recurrent nova RS~Ophiuchi (RS~Oph) erupted and was soon detected in optical, X-ray and GeV gamma-rays \citep{Munari2021arXiv210901101M,Munari2022arXiv220301378M,Page2022MNRAS.514.1557P,Cheung2022ApJ...935...44C}. Atmospheric Cherenkov telescopes MAGIC and H.E.S.S. successfully measured the time-resolved energy spectra, which reveals the temporal evolution of the maximum particle energy  \citep{magic2022NatAs...6..689A,hess2022Sci...376...77H}. Interestingly, the peak of the TeV gamma-ray light curve is delayed by two days relative to the peak of the GeV gamma-ray light curve \citep{hess2022Sci...376...77H}. After the peak, both TeV and GeV gamma-ray emissions decay smoothly in a power-law shape and the decay slopes are roughly consistent with each other.

The power-law decaying light curves support the external shock scenario, where shocks are produced via interactions between the nova ejecta and dense wind from RG. \citep{Zheng2022PhRvD.106j3011Z} interpreted the temporal decay behavior of the gamma-ray emission with an adiabatic external shock expanding in the red giant wind. 
The argument of adiabatic shock was confirmed by other independent modelings \citep{Diesing2023ApJ...947...70D,Sarkar2023ApJ...951...62D}. However, radiative loss may become more important at later times and an adiabatic shock may transition to a radiative shock \citep{Metzger2014MNRAS.442..713M,Chomiuk2021ARA&A..59..391C}. A general model of shocks including radiative loss is needed to understand the long-term behavior of the light curve.

The discovery of TeV gamma-rays in nova outbursts opened a new window to study the particle acceleration mechanism, especially the maximum energy of accelerated protons. It is found that a single power-law plus a sub-exponential cutoff for the proton spectrum, $dN/dE\propto E^{-\alpha}e^{-(E/E_{\rm max})^{0.5}}$, can roughly explain the gamma-ray spectra of the outburst of RS~Oph from GeV to TeV band \citep{hess2022Sci...376...77H,Zheng2022PhRvD.106j3011Z}. The more common cutoff shape, $dN/dE\propto E^{-\alpha}e^{-(E/E_{\rm max})}$, would, however, require some additional components to explain the GeV-TeV gamma-ray emission, such as multiple shocks \citep{Diesing2023ApJ...947...70D} or leptonic processes \citep{Sarkar2023ApJ...951...62D}.

In 2015, American Association of Variable Star Observers (AAVSO) found that the B-band light curve of T~CrB had a transition to a different state, whose curve shape in the B and V bands are similar to the pre-erupton phase around 1938. 
\cite{Munari2016NewA...47....7M} confirmed the state transition spectroscopically and called attention to this as
being similar to the transition in 1938.
Based on the similarity, \cite{Schaefer2019AAS...23412207S} predicted that the next eruption may occur in 2023.6$\pm$1.0. With follow-up observations, \cite{Luna2020ApJ...902L..14L} predicted that the eruption will occur in 2023–2026. Currently, the B-band light curve of T~CrB is in the pre-eruption dip, which predicts that the eruption date is in 2024 \citep{Schaefer2023ATel16107....1S}.

In Section \ref{model}, we first introduce our model for studying the shock evolution by including the radiative loss of the shock. Then, we apply this model to T~CrB in Section \ref{app} and study the observation signature for the transition from the adiabatic phase to the radiative phase. We then compute the GeV-TeV emission spectra and compare them with the sensitivity of MAGIC, VERITAS and LHAASO. 
We discuss the constraints on the particle acceleration physics that could be obtained with future TeV observations. A final conclusion and discussion are given in Section \ref{sec:con}.

\section{An External shock model for nova outbursts and application to T~CrB}
\label{model}
\subsection{External shock model}

The total energy of the shock includes the kinetic energy and the internal energy \citep[e.g.][]{Ostriker1988RvMP...60....1O,Padmanabhan2001thas.book.....P}. If the radiative loss is negligible, the  energy of the shock can be expressed by 
\begin{equation}
    \label{energy}
    E_{\rm sh,0}=\frac{1}{2}(M_{\rm ej}+m_{\rm sw})v^2_{\rm sh}+\frac{9}{32}m_{\rm sw}v^2_{\rm sh},
\end{equation}
where $M_{\rm ej}$ is the initial mass of the ejecta, $m_{\rm sw}$ is the mass of swept-up matter, $v_{\rm sh}$ is the velocity of the shock. The factor $9/32$ in the term of the internal energy is derived from the shock-jump conditions. 

As the nova expands into the ambient medium, new material is swept into the shock and the shock velocity gradually decreases after the mass of swept-up matter is comparable to that of the initial ejecta. If the initial radiative loss is small, the shock is adiabatic at early times. Its dynamics are identical to the Sedov phase of supernova remnants (SNRs). At later times, the radiative losses from the shock become significant and this stage is identical to the pressure-driven snowplow phase of SNRs.

Considering the thin-shell approximation where the nova ejecta $M_{\rm ej}$ and the swept-up matter $m_{\rm sw}$ are confined to a thin layer behind the shock, the energy loss rate of a fully radiative shock is (see Eq. 6.8 in \cite{Ostriker1988RvMP...60....1O}) 
\begin{equation}
    \label{lossrate}
    \frac{dE_{\rm sh}(t)}{dt}=-4\pi r^2_{\rm sh}\left(\frac{1}{2}\rho v^3_{\rm sh}\right),
\end{equation}
where $\rho$ is the density of ambient environment, and $\rho v^3_{\rm sh}/2$ is the energy flux density across the shock. In this stage, all of the energy of the swept-up material is radiated away. 
In reality, between the adiabatic phase and radiative phase, only a portion of energy flux $\epsilon_{\rm rad}\rho v^3_{\rm sh}/2$ radiates, where $\epsilon_{\rm rad}$ is defined as the radiation efficiency, which ranges from 0 to 1.  Therefore, the shock energy after considering the radiative loss is given by
\begin{equation}
    E_{\rm sh}(t)=\frac{1}{2}(M_{\rm ej}+m_{\rm sw})v^2_{\rm sh} + \frac{9}{32}(1-\epsilon_{\rm rad})m_{\rm sw}v^2_{\rm sh}.
\end{equation}
Considering the energy loss rate (Eq. \ref{lossrate}), we derive the dynamic equation of shock evolution, which is
\begin{equation}
    \label{dyn0}
    \frac{dv_{\rm sh}(t)}{dt}=-\frac{1}{2}\frac{(25+7\epsilon_{\rm rad})v_{\rm sh}(t)}{16M_{\rm ej}+(25-9\epsilon_{\rm rad})m_{\rm sw}(t)}\frac{dm_{\rm sw}(t)}{dt}.
\end{equation}
The increasing rate of swept-up mass is $dm_{\rm sw}/dt=4\pi\rho r^2_{\rm sh}v_{\rm sh}$. This dynamic equation can smoothly model the shock transition between the adiabatic phase and the radiative phase. It is difficult to obtain an analytical solution to the equation because the shock velocity $v_{\rm sh}(t)$ and swept-up $m_{\rm sw}(t)$ are both time-dependent. To analyse the self-similar solution of the shock, we re-write the Eq. (\ref{dyn0}) as
\begin{equation}
    \label{dyn1}
    \frac{dv_{\rm sh}(t)}{dm_{\rm sw}(t)}=-\frac{1}{2}\frac{(25+7\epsilon_{\rm rad})v_{\rm sh}(t)}{16M_{\rm ej}+(25-9\epsilon_{\rm rad})m_{\rm sw}(t)}.
\end{equation}

In the adiabatic phase, the radiation efficiency is zero ($\epsilon_{\rm rad}=0$), then the solution of  Eq. (\ref{dyn1}) is $v_{\rm sh}\propto(M_{\rm ej}+m_{\rm sw})^{-1/2}$, leading to $v_{\rm sh}\propto m^{-1/2}_{\rm sw}$ in the self-similar phase. For an ambient medium with a density $\rho\propto r^{-k}$, the shock velocity evolves as $v_{\rm sh}\propto r^{-(3-k)/2}_{\rm sh}$, resulting in $v_{\rm sh}\propto t^{-(3-k)/(5-k)}$ and $r_{\rm sh}\propto t^{2/(5-k)}$. In the fully radiative phase ($\epsilon_{\rm rad}=1$), the solution is $v_{\rm sh}\propto(M_{\rm ej}+m_{\rm sw})^{-1}$. So the shock velocity and radius evolve with time as $v_{\rm sh}\propto t^{-(3-k)/(4-k)}$ and $r_{\rm sh}\propto t^{1/(4-k)}$.
In a wind medium environment $\rho\propto r^{-2}$, the shock dynamics is described by $v_{\rm sh}\propto t^{-1/3}$ and $r_{\rm sh}\propto t^{2/3}$ for an adiabatic shock, while for a radiative shock,  $v_{\rm sh}\propto t^{-1/2}$ and $r_{\rm sh}\propto t^{1/2}$.  

We use the cooling timescale to calculate the radiation efficiency $\epsilon_{\rm rad}=t^{-1}_{\rm cool}/(t^{-1}_{\rm cool}+t^{-1}_{\rm ad})$, where $t_{\rm ad}=r_{\rm sh}/(2v_{\rm sh})$ is the dynamic timescale and $t_{\rm cool}$ is the radiative cooling timescale. In nova ejecta, the main cooling channels are free-free emission with $\Lambda_{\rm ff}=2\times10^{-24} T^{1/2}_{\rm sh,6} {\rm erg\,cm^{3}\,s^{-1}}$ and line emission with $\Lambda_{\rm line}=1.6\times10^{-22} T^{-0.7}_{\rm sh,6} {\rm erg\,cm^{3}\,s^{-1}}$ \citep{Schure2009A&A...508..751S}, where  $\Lambda$ is the cooling function.  The shock temperature is
\begin{equation}
    T_{\rm sh}=\frac{3\mu m_{\rm p}v^2_{\rm sh}}{16k_{\rm B}}=1.7\times10^{7} v^{2}_{\rm sh,8}{\rm K},
\end{equation}
where $k_{\rm B}$ is the Boltzmann constant, and $\mu=0.76$ is the mean molecular weight for nova ejecta \citep{Schwarz2007ApJ...657..453S}. Hereafter, we adopt the convention that subscript numbers $x$ indicate normalization by $10^x$ in cgs units. 

The transition of two cooling mechanisms occurs around $T_{\rm c}\sim5\times10^{7}$K, corresponding to the shock velocity $v_{\rm sh,c}=1.7\times10^{3} {\rm km s^{-1}}$.
The cooling time of the shock is $t_{\rm cool}=3k_{\rm B}T_{\rm sh}/2n_{\rm sh}(\Lambda_{\rm ff}+\Lambda_{\rm line})$, where $n_{\rm sh}=4n$ is the post-shock density. We use the accurate line emission cooling function given by plasma simulations (see Fig.1 in \cite{Schure2009A&A...508..751S}). 
For novae with high initial velocities ($v_0>v_{\rm sh,c}$), free-free emission is dominant in the early epoch, so the radiative efficiency is $\epsilon_{\rm rad}\approx t_{\rm ad}/t_{\rm cool}=0.05n_{9}r_{\rm sh,13}v^{-2}_{\rm sh,8.5}$. The shock is roughly adiabatic if the ambient density is small, $n\leq10^{9}{\rm cm^{-3}}$. At later times, as the shock decelerates, the shock will gradually transition to the radiative phase, which could be observed if the detectable time is sufficiently long.

\subsection{Probing shock dynamics with GeV observations}
\label{app}
There are only eleven recurrent novae discovered in our galaxy \citep{Schaefer2010ApJS..187..275S,Schaefer2022MNRAS.517.3864S}. Four of them have a red giant companion (V745 Sco,V3890 Sgr,RS~Oph and T~CrB). Among them, three novae in symbiotic systems (except for T~CrB)  have been detected in gamma-rays. These novae all have long orbital periods ($P>100$ days) and massive WD ($M_{\rm WD}\sim1.3M_{\odot}$) \citep{Belczynski1998MNRAS.296...77B,Brandi2009A&A...497..815B,Schaefer2010ApJS..187..275S,Shara2018ApJ...860..110S}, suggesting similarities in their ambient environments and explosion mechanisms. 

T~CrB is one of the brightest novae in the optical band. Previous outbursts that occurred in 1866 and 1946 have the peak apparent magnitude $m_{\rm v,peak}$=2.0 \citep{Schaefer2023MNRAS.524.3146S}, which is nearly ten times brighter than RS~Oph ($m_{\rm v,peak}$=4.8) \citep{Schaefer2010ApJS..187..275S,magic2022NatAs...6..689A}. This is due to the fact that the distance of T~CrB ($d$=887pc) is smaller than that of RS~Oph ($d$=2.45kpc) \citep{Rupen2008ApJ...688..559R,Bailer2021AJ....161..147B}. Their peak absolute magnitude $M_{\rm v,peak}$ are very close. This advantage would make the predicted GeV-TeV outburst brighter than that of RS~Oph, which will enable the current gamma-ray telescopes to observe its gamma-ray emissions for a longer time and with better precision. In this section, we will show how the GeV-TeV observations of T~CrB can be used to study the shock dynamics and probe particle acceleration physics.

T~CrB is embedded in the dense wind of a red giant, so the density far from the WD is approximately described by $\rho=Ar^{-2}$, where $A=\dot{M}/(4\pi v_{\rm w})=5\times 10^{12} A_{\star,0}\,{\rm g\,cm^{-1}}$.
We define $A_{\star}$ as the density corresponding to a mass loss rate of $\dot{M}=10^{-6}M_{\odot}\,\mathrm{yr}^{-1}$ and a wind velocity of $v_{\mathrm{w}}=10\,{\rm km\,s^{-1}}$.

For an initial adiabatic shock, the self-similar solution is $v_{\rm sh}\propto t^{-1/3}$ and $r_{\rm sh}\propto t^{2/3}$ after the deceleration. At the beginning, free-free emission is the dominated  cooling channel of the shock,  $t_{\rm cool}\propto T_{\rm sh}/\rho\Lambda_{\rm ff}$. Noting that $\Lambda_{\rm ff}\propto T_{\rm sh}$ and $T_{\rm sh}\propto v^2_{\rm sh}$, we have $t_{\rm cool}\propto v_{\rm sh}r^2_{\rm sh}\propto t$. Because the dynamic timescale also increases with time $t_{\rm ad}\approx r_{\rm sh}/v_{\rm sh}\propto t$, the cooling efficiency  is constant $\epsilon_{\rm rad}(t_{\rm dec})=0.27A_{\star,0}^{2} M^{-1}_{\rm ej,-6} v^{-2}_{\rm 0,8.7}$ in this stage, where the reference value of $v_0$ is $5000{\rm kms^{-1}}$.

The adiabatic phase ends at $t_{\rm c}$ when the line emission dominates the cooling. Adopting the deceleration timescale $t_{\rm dec}={M_{\rm ej}}/({4\pi A v_0})$ as the beginning of the self similar evolution $v_{\rm sh}\propto t^{-1/3}$, we obtain the transit time
\begin{equation}
    t_{\rm c}=t_{\rm dec}\left(\frac{v_0}{v_{\rm sh,c}}\right)^3=19\,A_{\star,0}^{-1} M_{\rm ej,-6} v^2_{\rm 0,8.7} \,{\rm day}.
\end{equation}

\begin{table}
    \centering
    \caption{Physical properties of T~CrB and RS~Oph.}
    \label{tab:my_label}
    \begin{threeparttable}
    \begin{tabular}{ccc}
    \hline
    \hline
      & T~CrB  & RS~Oph  \\
    \hline
     $m_{\rm V,peak}$ &   2.0\tnote{1} &  4.8\tnote{2} \\
     $M_{\rm V,peak}$ &  -7.7 & -7.2 \\
     $P_{\rm orb}$ (days) & 227.27$\pm$0.001\tnote{2} & 453.6$\pm$0.4\tnote{2} \\
     Eccentricity & 0.012$\pm$0.005\tnote{3} & 0.04$\pm$0.03\tnote{4} \\
     Inclination angle ($^{\circ}$) & 60$\pm$5\tnote{3} & 50.5$\pm$1.5\tnote{4} \\
     $M_{\rm WD}$ ($M_{\odot}$) & 1.2$\pm$0.2\tnote{3} & 1.3$\pm$0.1\tnote{4} \\
     $M_{\rm RG}$ ($M_{\odot}$) & 0.7$\pm$0.2\tnote{3} & 0.74$\pm$0.06\tnote{4} \\
     $\dot{M}_{\rm acc}$ ($10^{-8}M_{\odot}{\rm yr^{-1}}$) & 2.1\tnote{5} & 7.2\tnote{5} \\
    \hline
    \hline
    \end{tabular}
    \begin{tablenotes}
        \footnotesize
        \item[1] {\cite{Schaefer2023MNRAS.524.3146S}}
        \item[2] {\cite{Schaefer2010ApJS..187..275S}}
        \item[3] {\cite{Belczynski1998MNRAS.296...77B}}
        \item[4] {\cite{Brandi2009A&A...497..815B}}
        \item[5] {\cite{Shara2018ApJ...860..110S}}
    \end{tablenotes}
    \end{threeparttable}
\end{table}

After transiting to the line emission dominant cooling phase, the cooling time is  $t_{\rm cool}\propto T^{1.7}_{\rm sh}/\rho\propto v_{\rm sh}^{3.4}r_{\rm sh}^2\propto t^{0.2}$. During this phase, the radiative efficiency increases with time $\epsilon_{\rm rad}\propto t^{4/5}$, and thus the shock will become fully radiative at a critic time $t_{\rm r}$, which is given by
\begin{equation}
    \label{tr}
    t_{\rm r}=t_{\rm c}\epsilon^{-5/4}_{\rm rad}(t_{\rm dec})=97\, A_{\star,0}^{-3.5} M^{2.25}_{\rm ej,-6}v^{4.5}_{\rm 0,8.7} \,{\rm day}.
\end{equation}

\begin{figure}
    \centering
    \includegraphics[width=1.0\columnwidth]{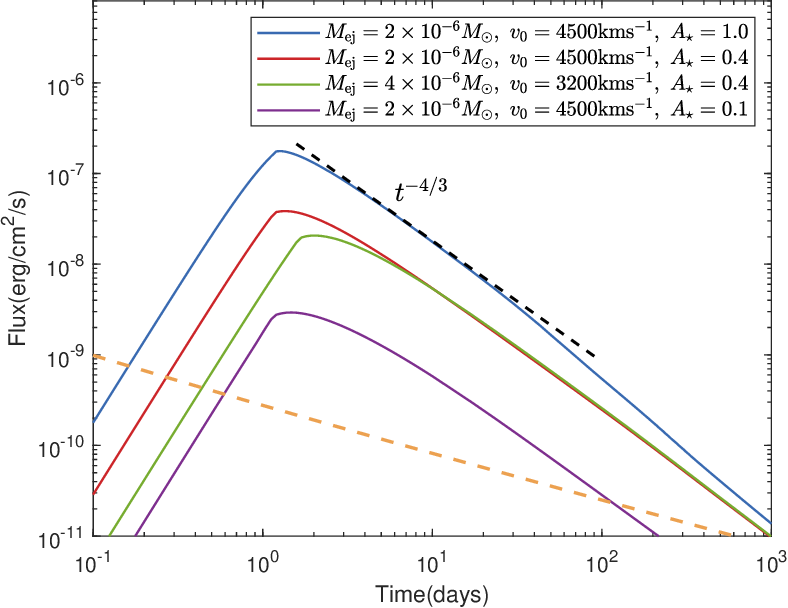}
    \caption{The gamma-ray $(0.1-10 {\rm GeV})$ light curves of an outburst from T~CrB. The blue, red and purple lines represent gamma-ray light curves produced in different environments $A_{\star}=1$, $A_{\star}=0.4$ and $A_{\star}=0.1$, respectively. Other parameters for these three lines are $M_{\rm ej}=2\times10^{-6}M_{\odot}$ and $v_0=4500{\rm kms^{-1}}$. The green line shows the gamma-ray produced by $A_{\star}=0.4$, $M_{\rm ej}=4\times10^{-6}M_{\odot}$ and $v_0=3200{\rm kms^{-1}}$.
    The black dashed line exhibits a light curve decaying as $F_{\gamma}\propto t^{-4/3}$.  The orange dashed line is the flux sensitivity curve of Fermi-LAT (see Appendix A for details).  
    }
    \label{fig:lat}
\end{figure}

The gamma-ray emission from nova outbursts are likely produced by the hadronic process \citep[e.g.][]{Li2017NatAs...1..697L,Chomiuk2021ARA&A..59..391C,magic2022NatAs...6..689A,hess2022Sci...376...77H}, where relativistic protons accelerated by nova shocks collide with circum-stellar wind (i.e., $pp$ interaction), producing pions that decay into gamma-rays.
In the adiabatic phase, gamma-ray emissions decline as $F_{\rm \gamma}\propto t^{-4/3}$. The decline  gradually transits to $F_{\rm \gamma}\propto t^{-3/2}$ when the radiative loss is important \citep{Zheng2022PhRvD.106j3011Z}. Fermi-LAT detected significant GeV signals from RS~Oph only in the first month of its outburst. The decay slope agrees  with $F_{\rm \gamma}\propto t^{-4/3}$ \citep{hess2022Sci...376...77H,Zheng2022PhRvD.106j3011Z}. At later times, the source becomes undetectable, so the transition between the radiative phase and the adiabatic phase could not be tested.

Since T~CrB is an order of magnitude brighter than RS~Oph,  the detectable time by Fermi-LAT could potentially last for up to 170 days, assuming the same decaying slope of $t^{-4/3}$. The extended observation period could probably enable us to capture the transition of shock evolution if the wind density is not too small. We compute the gamma-ray emission (0.1-10 GeV) light curve from a realistic shock, which takes into account the time-dependent radiative loss, for the predicted outburst of T~CrB.
We assume that, similar to RS~Oph, the density profile is a broken power-law shape where the density is nearly a constant at small radii while at large radii the density approaches the wind profile $\rho\propto r^{-2}$ \citep{Zheng2022PhRvD.106j3011Z}. The transition occurs at $r_{\rm w}=3a$ where $a\approx1.5\times10^{13}$cm is the semi-major axis for T~CrB \citep{Belczynski1998MNRAS.296...77B}. The transition radius is consistent with the three dimensional simulation of RS~Oph \citep{Walder2008A&A...484L...9W}.

The starting time of the fully radiative phase is sensitive to the nova kinetic energy $E_{\rm k}=M_{\rm ej}v^2_0/2$ and the ambient environment $A_{\star}$.
 The dynamical parameters (i.e. $M_{\rm ej}$ and $v_0$) of T~CrB and RS~Oph are likely to be similar because the WD masses in these two systems are close (see Table 1). 
 
 Nova outbursts are triggered by the thermonuclear runaway mechanism,  the energetics of which are mainly determined by the properties of the WD \citep{Starrfield2016PASP..128e1001S}. 
 Numerical simulations showed that for a WD near the Chandrasekhar limit ($M_{\rm WD}=1.25-1.35 M_{\odot}$), the outburst will erupt ejecta $M_{\rm ej}\approx10^{-6}-10^{-5} M_{\odot}$ with maximum velocity $v_{\rm max}\approx3000-6000 {\rm kms^{-1}}$ \citep{Starrfield2009ApJ...692.1532S}. In the 2021 outburst of RS~Oph, the initial velocity and ejecta mass were $v_0\approx4500{\rm kms^{-1}}$ and $M_{\rm ej}\approx2\times10^{-6}M_{\odot}$ \citep{magic2022NatAs...6..689A,Zheng2022PhRvD.106j3011Z}, which is consistent with theoretical expectations.

Furthermore, \cite{Hachisu2001ApJ...558..323H} modeled the optical light curves of RS~Oph and T~CrB and suggested two outbursts both originate from a very massive WD ($M_{\rm WD}\approx1.35M_{\odot}$) and their ejecta masses are $M_{\rm ej}\approx2\times10^{-6}M_{\odot}$. Therefore, we apply the same ejecta mass $M_{\rm ej}=2\times10^{-6}M_\odot$ and initial velocity $v_0\approx4500{\rm kms^{-1}}$ to T~CrB as reference parameters. The ejecta mass may also be influenced by the accretion rate of the WD \citep[e.g.][]{Chen2019MNRAS.490.1678C}. Since the accretion rate of these two systems are comparable \citep{Shara2018ApJ...860..110S}, we estimate the difference in the ejecta mass is within a factor of two.

It's challenging to constrain the wind density $A_{\star}$ with our current knowledge. Therefore, we plot GeV gamma-ray light curves with different $A_{\star}$ values in Fig .\ref{fig:lat}. 
The red and green lines represent GeV gamma-rays produced by shocks with different initial velocities. The peak time of GeV light curve is $t^{\rm GeV}_{\rm peak}=\min\left\{t_{\rm w},t_{\rm dec}\right\}$, where $t_{\rm w}=r_{\rm w}/v_0$ is the time that shock enters wind environment. 
After the shock decelerates, the gamma-ray light curves are determined by the kinetic energy and wind density. The green and red lines converge after $\sim$5 days because they have the same kinetic energy.
We also discover that for adiabatic shocks in the self-similar phase, the gamma-ray flux follows $F_{\gamma}\propto E^{1/3}_{\rm k}A^{5/3}_{\star}$. The GeV light curves with $A_{\star}=1$ (blue line) deviates from $F_{\gamma}\propto t^{-4/3}$ because of radiative loss. The shock enters the snowplow phase at $\sim220$ days. We compare the sensitivity curve of Fermi-LAT and theoretical models in Figure \ref{fig:lat} (see Appendix A for details) and find that T CrB is detectable until $\sim$ 1000 days. Therefore, Fermi-LAT could probably capture the transition of shock phases if $A_{\star}\geq1$.



\begin{figure*}[ht]
    \centering
    \includegraphics[width=0.9\columnwidth]{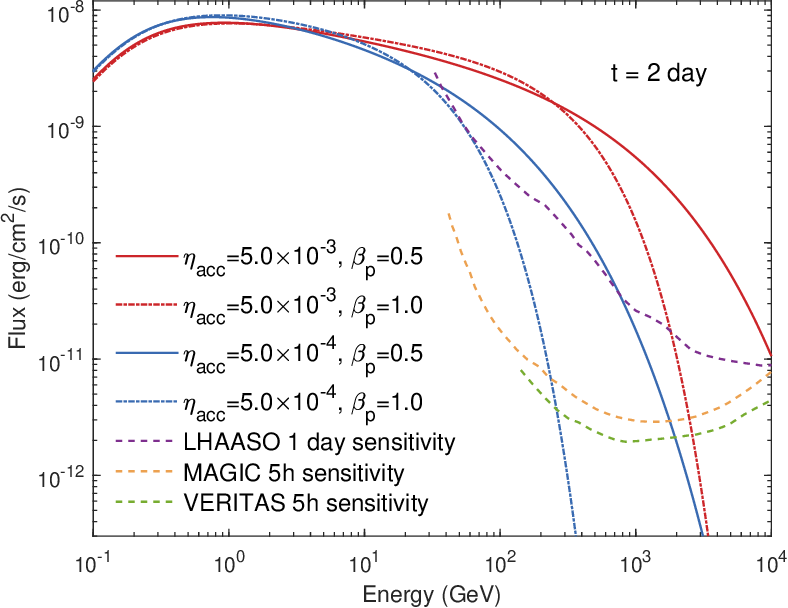}
    \includegraphics[width=0.9\columnwidth]{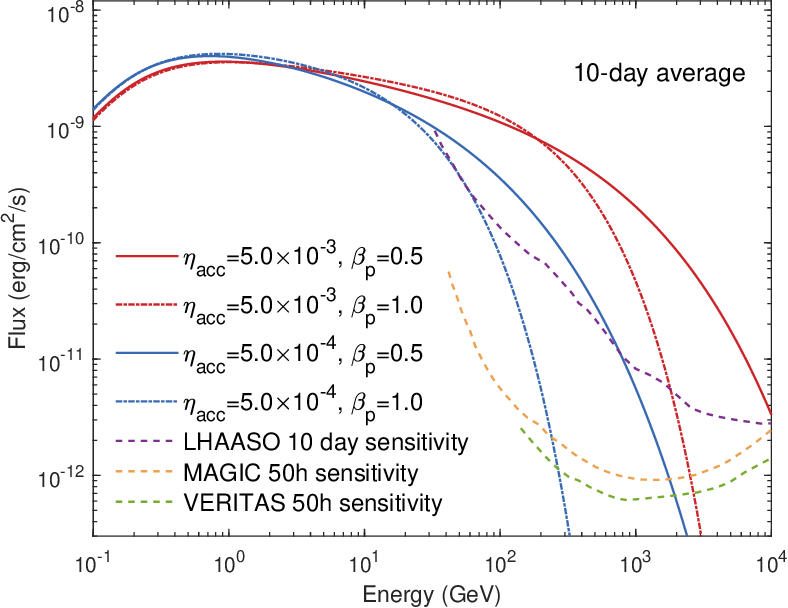}
    \caption{
    Comparison between the expected gamma-ray spectra of T~CrB outburst and sensitivity curves of VHE gamma-ray telescopes. {\textit{Left panel}:} Gamma-ray spectra of T~CrB at $t=2$ day. The red and blue lines represent the gamma-ray spectra with acceleration efficiency $\eta_{\rm acc}=5\times10^{-3}$ and $\eta_{\rm acc}=5\times10^{-4}$, respectively. The solid lines correspond to the cutoff index $\beta_{\rm p}=0.5$ and the dash-dotted lines correspond to the cutoff index $\beta_{\rm p}=1.0$.
    The purple dashed line represents the sensitivity curve of LHAASO over the duration of one day while the orange and green dashed lines represent the 5h sensitivity curves of MAGIC and VERITAS, respectively. {\textit{Right panel}:} The expected averaged gamma-ray spectra of T~CrB outburst in the first ten days. The symbols follow the same convention as in the right panel. The purple dashed line is the sensitivity curve of LHAASO in ten days and the orange and green dashed lines are 50h sensitivity curves of MAGIC and VERITAS, respectively.
    }
    \label{fig:wcda}
\end{figure*}

In the calculation, we assume a power-law proton spectrum with a high-energy cutoff, which is
\begin{equation}
    \frac{\mathrm{d} N}{\mathrm{d} E_{\rm p}}=C_{\mathrm{E}}E^{-\alpha_{\rm p}}_{\rm p}{\mathrm{exp} \left[-\left(\frac{E_{\rm p}}{E_{\rm max}}\right)^{\beta_{\rm p}}\right]}.
\end{equation}
We use the equipartition factor $\epsilon_{\rm p}=0.2$ representing a fraction of energy is transferred to accelerated protons and the normalization factor $C_{\rm E}$ is given by $\int^{E_{\rm max}}_{E_{\rm min}} E_{\rm p}({\mathrm{d}N_{\rm p}}/{\mathrm{d}E_{\rm p}})\mathrm{d}E_{\rm p}=\epsilon_{\rm p}(1/2+9(1-\epsilon_{\rm rad})/32)m_{\rm sw}v^2_{\rm sh} $. We adopt $\alpha_{\rm p}=2.2$ and $E_{\rm min}=m_{\rm p}c^2$ for non-thermal protons.  To avoid the influence from the unknown maximum proton energy, we fix the $E_{\rm max}=100$GeV in the computation of Fig. \ref{fig:lat}. Since high-energy cutoff does not affect the 0.1-10 GeV emission, the value of $\beta_{\rm p}$ is unimportant for GeV emission.

\section{Probing particle acceleration physics with TeV observations}
T~CrB is located in the northern sky ( R.A. = $239.9^{\circ}$ Dec.=$+25.9^{\circ}$), offering favorable conditions for observations by  TeV telescopes in the northern hemisphere (e.g. MAGIC, VERITAS, LHAASO). Imaging Atmospheric Cherenkov Telescopes (IACTs) have better sensitivity in the TeV band for short-period observations, but they could be affected by moonlight \citep{Griffin2015ICRC...34..989G,MAGIC2016APh....72...61A}. 
Acting as a wide-aperture and high-duty-cycle gamma-ray detector in VHE gamma-rays, LHAASO has advantages of capturing the early stage evolution and continuously observing the long-term evolution of T~CrB. The latitude of LHAASO site is about $+29^{\circ}$, which means the smallest zenith angle for T~CrB is nearly zero, so the observation time in low zenith angle ($\theta_{\rm z}<30^{\circ}$) is more than four hours in one day.  The sensitivity of LHAASO outperforms that of HAWC in the 0.3-10 TeV range \citep{hawc2017ApJ...843...39A}. For brevity, we only choose MAGIC, VERITAS and LHAASO to compare their sensitivity curves with the theoretical models.


The detectability of TeV gamma-ray emission is sensitive to the cutoff index $\beta_{\rm p}$ (see Eq. 9) in the proton spectrum. Previous models of the nova outbursts used different cutoff index $\beta_{\rm p}$ in the spectrum of injected protons \citep{hess2022Sci...376...77H,Zheng2022PhRvD.106j3011Z,Diesing2023ApJ...947...70D}. 
The cutoff index $\beta_{\rm p}$ is determined by the particle transport mechanism inside the shock. \cite{Caprioli2009MNRAS.396.2065C} derived a stationary solution for the transportation equation, which shows that the cutoff index $\beta_{\rm p}$ results from the diffusion coefficient of particles, $D(E)\propto E^{\beta_{\rm p}}$ (see their Eq. 4). The exponential cutoff $\beta_{\rm p}=1$ corresponds to the Bohm diffusion $D(E)\propto E$ while the sub-exponential cutoff $\beta_{\rm p}=0.5$ suggests a different diffusion process in the shock. Therefore, one can test the different diffusion processes in the shock if a precise gamma-ray spectrum near the cutoff can be measured.


TeV gamma-ray spectra are also sensitive to the maximum accelerated proton energy.  By equating the Lamour radius  with the shock radius, one obtains (the Hillas condition, details see \cite{Gaisser2016crpp.book.....G}) 
\begin{equation}
    E_{\rm max}\approx\frac{\eta_{\rm acc}v_{\rm sh} eBr_{\rm sh}}{c},
\end{equation}
where $\eta_{\rm acc}$ is the acceleration efficiency, $c$ is the speed of light and $B$ is the magnetic field in the post-shocked medium. Assuming that the magnetic energy density behind the shock is proportional to the shock internal energy density, we get the magnetic field  $B=\sqrt{8\pi\varepsilon_{\rm B}\rho v^2_{\rm sh}}=1.9\,{\rm G}A^{1/2}_{\star,-0.4}\varepsilon^{1/2}_{\rm B,-2.5}r^{-1}_{\rm sh,14}v_{\rm sh,8.7}$, where $\varepsilon_{\rm B}$ is the equipartition factor of magnetic field. Observations for SNRs suggest that $\varepsilon_{\rm B}$ is within $10^{-3}-10^{-2}$ (see Table 3 in \cite{Helder2012SSRv..173..369H}), so we use the median value $\varepsilon_{\rm B}=3\times10^{-3}$ in our calculation.


The maximum accelerated energy is then $E_{\rm max}=9.7\,{\rm TeV}\eta_{\rm acc,-2}A^{1/2}_{\star,-0.4}\epsilon^{1/2}_{\rm B,-2.5}v^2_{\rm sh,8.7}$.
Here we use the parameter values ($M_{\rm ej}=2\times10^{-6}M_\odot$, $v_0=4500\,{\rm km \,s^{-1}}$, and $A_{\star}=0.4$) of RS~Oph as the reference values to calculate the gamma-ray emissions of T~CrB. For these parameter values, TeV emissions peak at $t\approx2$ days, and the gamma-ray spectra for two different acceleration efficiencies ($\eta_{\rm acc}=5\times 10^{-4}$ and $\eta_{\rm acc}=5\times 10^{-3}$)  at the peak time are shown in the right panel of Fig. \ref{fig:wcda}. The sensitivity curves of MAGIC and VERITAS are taken from \cite{magic2016APh....72...76A} and \cite{veritasweb}. We also show the spectra of gamma-rays with two different cutoff indices ($\beta_{\rm p}=0.5$ and $\beta_{\rm p}=1.0$).

In the case of $\beta_{\rm p}=0.5$ and $\eta_{\rm acc}=5\times10^{-4}$, the maximum proton energy is nearly 300 GeV,  which is close to the inferred maximum proton energy for RS~Oph from TeV observations \citep{Zheng2022PhRvD.106j3011Z}. The flux at 300 GeV is comparable to that of the Crab nebula, $F_{\gamma}\approx10^{-10}{\rm erg}{\rm cm}^{-2}{\rm s}^{-1}$, which means IACTs such as VERITAS can collect significant signals within about 5 minutes \citep{veritasweb}. For an exponential cutoff spectrum with $\beta_{\rm p}=1$, IACTs can still detect enough signals in 100-300 GeV if $\eta_{\rm acc}\geq5\times10^{-4}$. The observation energy band covers the entire cutoff shape of the photon spectra, which is crucial for identifying the cutoff index $\beta_{\rm p}$ of the proton spectra.

The number of  photons  detected by LHAASO is estimated to be
\begin{equation}
    N_{\rm s}=\iint \frac{dN}{dE_{\gamma}}A_{\rm eff}(E_{\gamma},\theta_{\rm z})dE_{\gamma}dt,
\end{equation}
where $\theta_{\rm z}$ is the zenith angle of the source, $A_{\rm eff}(E_{\gamma},\theta_{\rm z})$ is the effective area of the detector, depending on the zenith angle and photon energies. We used the effective area of WCDA and KM2A provided by  \cite{lhaasoWhitePaper} and \cite{km2a2021ChPhC..45b5002A}, respectively.
In the RS~Oph-like case ($\beta_{\rm p}=0.5$ and $\eta_{\rm acc}=5\times10^{-4}$), We expect there are more than 1800 photons with energies greater than 300 GeV detected by LHAASO-WCDA from T~CrB outburst during the peak day, which would result in a significant detection of $\sim7.5\sigma$. However, for an exponential cutoff with $\beta_{\rm p}=1.0$ and the same acceleration efficiency $\eta_{\rm acc}=5\times10^{-4}$, the gamma-ray signal is only marginal for detection because of the sharp cutoff of the photon spectra around 30 GeV.  LHAASO-WCDA detection would require $\eta_{\rm acc}\geq10^{-3}$ in the $\beta_{\rm p}=1.0$ case for a significant detection.
We also show the average spectra in the first ten days in the left panel of Fig. \ref{fig:wcda}. During this longer period, photons collected by WCDA are expected to be more than 7000 in the RS~Oph-like case. Conclusions about the detectability for IACTs and LHAASO remain the same. The gamma-ray flux in early 10 days roughly follows the scaling relation $F_{\gamma}\propto A^{5/3}_{\star}$ ,which can be used to estimate the significance if the ambient environment of T~CrB is different with RS~Oph.

\begin{figure}
    \centering
    \includegraphics[width=1.0\columnwidth]{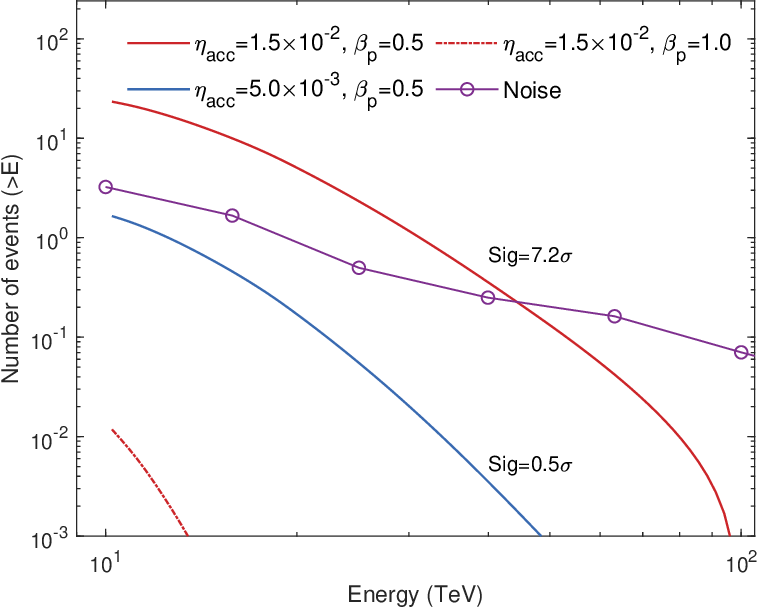}
    \caption{
    The expected detected photon numbers of T~CrB outburst by LHAASO-KM2A as a function of the photon energy at $t=2$ day. The purple open circles denote the noise of cosmic ray background events $\sqrt{N_{\rm b}(\Psi_{70})}$. The red and blue lines represent $71.5\%$ signal photon numbers with acceleration efficiencies $\eta_{\rm acc}=1.5\times10^{-2}$ and $\eta_{\rm acc}=5\times10^{-3}$, respectively. The solid lines indicate the cutoff index $\beta_{\rm p}=0.5$ and the dot-dashed line indicates the cutoff index $\beta_{\rm p}=1.0$. We show the significance level of the signal for $\beta_{\rm p}=0.5$ cases. The line of $\eta_{\rm acc}=5\times10^{-3}$ and $\beta_{\rm p}=1.0$ case is not shown because the photon number is smaller than $10^{-3}$.
    }
    \label{fig:km2a}
\end{figure}



Above 10 TeV,  LHAASO-KM2A is more sensitive than IACTs. The cosmic-ray background of LHAASO-KM2A in a cone of $1^{\circ}$ was given in \citet{lhaaso2021Sci...373..425L}. Because the Point Spread Function (PSF) of LHAASO-KM2A is  $\delta\Psi\approx0.25^{\circ}$, the CR background could be reduced by an additional factor of $(1^{\circ}/\delta\Psi)=16$. However, downsizing the aperture also reduces the strength of signals. To maximize the signal to noise ratio, we use the aperture $\Psi_{70}=1.58\delta\Psi$ in the calculations, which covers $71.5\%$ signals. The significance of the signal is calculated by $Sig=0.715N_{\rm s}/\sqrt{N_{\rm b}(\Psi_{70})}$.
We found that the signal  is lower than the noise when the zenith angle of the source is larger than $30^{\circ}$ because the effective area of LHAASO-KM2A significantly decreases above this angle. Hence, the effective observing time for T~CrB is only $\sim$4 hours per day. 

We calculate the detectability of T~CrB outburst at $t=2$ day by LHAASO-KM2A for various values $\eta_{\rm acc}$ and $\beta_{\rm p}$, which is shown in Fig.\ref{fig:km2a}. 
The three lines represent the photon numbers greater than a specific energy $N(>E)$. The background events above 10 TeV are approximately $N_{\rm b}(\Psi_{70})\approx$ 11 events within 4 hours. To reach the threshold of $5\sigma$ significance ($Sig\geq5$), the signal photons should be greater than $N_{\rm s}\geq23$, which requires the acceleration efficiency to be $\eta_{\rm acc}\geq0.015$ when $\beta_{\rm p}=0.5$, corresponding to the maximum proton energy $E_{\rm max}\approx10$ TeV. For an exponential cutoff spectrum with $\beta_{\rm p}=1$,  a very high acceleration efficiency $\eta_{\rm acc}\gtrsim0.1$ would be required for a significant detection by LHAASO-KM2A. For different environments, new constraints can be obtained by adopting the scaling relation $E_{\rm max}\propto\eta_{\rm acc}A^{1/2}_{\star}$ and $F_{\gamma}\propto A^{5/3}_{\star}$.

\section{Conclusions and Discussions}
\label{sec:con} 
We proposed a dynamic model for nova external shocks that includes the radiative loss of the shock. This model produces a smooth transition between the adiabatic phase and radiative phase for the nova shocks. The line emission plays an important role in the late shock evolution.
Therefore, our model predicts the shock transitions to a radiative shock at $97\, A_{\star,0}^{-3.5} M^{2.25}_{\rm ej,-6}v^{4.5}_{\rm 0,8.7}$ days, where $M_{\rm ej}=10^{-6}M_\odot$, $v_0=5000{\rm kms^{-1}}$ and $A_{\star}=1$ are reference parameter values.
The phase transition of the shock manifests itself in the decay slope of the GeV light curve, changing from $t^{-4/3}$ to $t^{-3/2}$, which may be observed in the upcoming outburst of the recurrent nova T~CrB if the wind environment $A_{\star}\geq1$.

The density of the ambient environment around T~CrB is not well-known. 
Although companion stars of RS~Oph and T~CrB are both M type red giants, the different orbital periods could lead to different ambient environments. The flux of gamma-ray emission is sensitive to the ambient density $F_{\gamma}\propto A^{5/3}_{\star}$ because it affects both the shock dynamics and the $pp$ interaction efficiency. 

As a bright nova, T~CrB offers a unique opportunity for studying the particle acceleration mechanism by measuring the spectra of VHE gamma-rays. The cutoff shape of the photon spectra reflects the cutoff index $\beta_{\rm p}$ of the proton spectra (see Eq. 9).
We calculate the gamma-ray flux by using our shock dynamic model and find that the peak flux at 300 GeV is comparable to that of the Crab Nebula, $F_{\gamma}\approx10^{-10}{\rm erg}{\rm~cm}^{-2}{\rm~s}^{-1}$, if the parameters of T~CrB are close to those of RS~Oph (i.e., $\beta_{\rm p}=0.5$ and $\eta_{\rm acc}=5\times10^{-4}$). The VHE emissions are expected to be detectable by IACTs and LHAASO, as shown in Fig. \ref{fig:wcda}.  For an exponential cutoff proton spectrum with $\beta_{\rm p}=1.0$, IACTs can still detect T~CrB around 100 GeV in the case of $\eta_{\rm acc}\geq5\times10^{-4} A^{-1/2}_{\star,-0.4}$, while LHAASO-WCDA requires $\eta_{\rm acc}\geq10^{-3}A^{-1/2}_{\star,-0.4}$ for a significant detection. 
Above 10 TeV, T~CrB could be detected by LHAASO-KM2A at a significance level of 5$\sigma$. In the case of $\beta_{\rm p}=0.5$ and $\eta_{\rm acc}\geq0.015A^{-1/2}_{\star,-0.4}$. For an exponential cutoff $\beta_{\rm p}=1.0$, the required acceleration efficiency for the detection is $\eta_{\rm acc}\gtrsim0.1A^{-1/2}_{\star,-0.4}$, which seems to be too extreme. Thus, measurements of VHE gamma-rays from the upcoming outburst of T~CrB could constrain the cutoff index in the proton spectrum and the acceleration efficiency, and hence constrain the particle acceleration mechanism in nova shocks.

We assumed $\varepsilon_{\rm B}=3\times10^{-3}$ in our calculation as it is the medium value obtained from SNR observations \citep{Helder2012SSRv..173..369H}. Noting that some SNRs have relatively high magnetic fields with $\varepsilon_{\rm B}=0.03$ \citep[e.g. SN 1993J,][]{Helder2012SSRv..173..369H}, the aforementioned constraints on $\eta_{\rm acc}$ could be relaxed by a factor of three.

Since novae are scaled-down supernovae, constraints on the proton spectra and the maximum energy in novae are important for us to test whether young supernova remnants are PeV cosmic-ray accelerators (PeVatrons). Noting that novae in symbiotic systems have similar wind environment to that in core collapsed supernovae (CCSNe), novae are particularly useful to study particle acceleration in the remnants of CCSN.  For a mass loss rate  $10^{-5}M_{\odot} {\rm yr}^{-1}$ from the progenitor of CCSN \citep{Vink2020pesr.book.....V}, the maximum proton energy is $E_{\rm max}=200\,{\rm TeV}\eta_{\rm acc,-2}A^{1/2}_{\star,1}\epsilon^{1/2}_{\rm B,-2.5}v^2_{\rm sh,9}$. A PeVatron would require that the acceleration efficiency is $\eta_{\rm acc}\geq0.05A^{-1/2}_{\star,1}\epsilon^{-1/2}_{\rm B,-2.5}v^{-2}_{\rm sh,9}$.

\section*{Acknowledgements}
The authors thank Yun-Lang Guo, Zhen Cao, Yong Huang and Shi-Cong Hu for useful discussions. The work is supported by the NSFC under grants Nos. 12333006, 12121003, 12393852, 12203022.

\appendix

\section{Fermi LAT sensitivity as a function of observation time}

We use the open-source Python analysis package \textit{FermiPy} \citep{2017ICRC...35..824W} to calculate the LAT flux sensitivity.
The \textit{Sensitivity Tools} in \textit{FermiPy} calculates the LAT flux threshold for a gamma-ray source in bins of energy (differential sensitivity) and integrated over the full LAT energy range (integral sensitivity). In this work, we use a livetime cube generated from real data and spacecraft file for $\sim16$ yrs of observation, and the Pass 8 instrument response function (IRF) $"P8R3\_SOURCE\_V3"$ is used. We assume a power-law spectrum for the point-like source, and set a TS threshold of 25 and a minimum of 10 excess counts. The orange dashed line in Fig. \ref{fig:lat} shows the sensitivity which integrated over 0.1--10 GeV energy band with a photon index of 2.5 in different time.


\bibliographystyle{elsarticle-harv}
\bibliography{reference.bib}






\end{document}